____________________________________________________________________

\documentstyle[12pt]{article}

\newcommand{\EQ}{\begin{equation}}
\newcommand{\EN}{\end{equation}}

\begin{document}

\topmargin 0pt
\oddsidemargin 5mm

\renewcommand{\thefootnote}{\fnsymbol{footnote}}

\newpage
\setcounter{page}{0}
\begin{titlepage}
\begin{flushright}
PITT-93-07 \\
May 1993
\end{flushright}
\begin{center}
{\large ON THE RELATION BETWEEN  \\
 EUCLIDEAN AND REAL TIME CALCULATIONS \\
OF GREEN FUNCTIONS AT FINITE TEMPERATURE} \\
\vspace{1cm}
{\large Alexander Bochkarev \footnote{On leave of absence from INR,
Russian Academy of Science, Moscow 117312, Russia} \\
{\em Department of Physics and Astronomy, \\
University of Pittsburgh, Pittsburgh, PA 15260} }
\end{center}
\vspace{1cm}

\begin{abstract}

We find a relation between the semiclassical approximation of the
temperature
(Matsubara) 2-point correlator and the corresponding classical Green
function
in real time at finite temperature. The anharmonic oscillator at finite
temperature
is used to illustrate our statement, which is however of rather general
origin.

\end{abstract}
\end{titlepage}

\newpage

The real time Green functions of a system at finite temperature do not
allow for the
conventional Feynman diagram technique. They are however related to the
so called {\em temperature Green fuctions} via analytical continuation.
The temperature Green functions are in the Matsubara \cite{mats}
representation,
which has a usual functional intergral expression. Then one is naturally
interested in the semiclassical approximations for the Matsubara functional
integrals as soon
as nonperturbative effects are in question. While the semiclassical
expansion is
certainly a powerful way to calculate energy levels, it is not apriori
clear
whether this approximation (developed in the Euclidean domain) remains
an efficient
expansion after the analytical continuation to Minkowski space, as far
as time
dependent Green functions are concerned.
The question is equally important for the quantum field theory at zero
temperature.
The instanton induced Green functions in the Electroweak model are
presently  a
subject of intense
debate \cite{mat}. They are computed by means of the semiclassical
expansion in the
Euclidean space and then analytically transformed into the real-time
\cite{rw}.
In \cite{dk} a disagreement was reported
between the classical ( $\hbar \rightarrow 0$ ) limit of the analytically
continued
leading semiclassical expression for the 2-point Matsubara correlator of
 the anharmonic oscillator at finite temperature and the classical 2-point
real time
Green function of the same theory. We clarify the nature of that
discrepancy which proves to be of rather general origin.

Let $G_{\beta, \hbar}(\tau)$ be a 2-point Matsubara correlator  :
\EQ
G_{\beta, \hbar}(\tau) \; \; = \; \; {\cal Z}^{-1} \; {\cal T} \,
\{ x(\tau) \, x(0) \}                \label{G}
\EN
where ${\cal T}$ stands for time-ordering, $\beta$ is inverse temperature
and the operator $x(\tau)$ is:
$x(\tau) \;=\; \exp(- \tau H / \hbar) \, x(0) \, \exp(\tau H / \hbar)$.
The Fourier transform of $G_{\beta, \hbar}(\tau)$ constitutes the same
analytic
function as the Fourier transform of the corresponding retarded and
advanced real time Green functions.
We are however interested in the classical correlation function of $x(t)$
which
may be obtained from (\ref{G}) by means of the analytical continuation
$\tau \,=\, i t$ in the limit  $\hbar \rightarrow 0$. We first compute
(\ref{G})
semiclassically , perform the limit  $\hbar \rightarrow 0$, do
analytical continuation and then compare the result with the direct
evaluation of the classical correlator.

The partition function  ${\cal Z}$ is the following Matsubara functional
integral:
\EQ
{\cal Z} \;=\;  \int_{x(0) = x(\beta \hbar)} {\cal D}x(\tau) \;
\exp \left[ - \frac{1}{\hbar}
\int_{0}^{\beta \hbar} d\tau\,\left( \frac{\dot{x}^{2}}{2} \,+\, V(x)
\right)\right]
                            \label{Z}
\EN
The integration over the end point $x(0) = x(\beta \hbar) = x$ may be
factorized:
\EQ
{\cal Z} \;=\; \int dx \; \int_{x(0) = x(\beta \hbar) = x} {\cal D}x(\tau)
\; \exp \left[ - {\cal S}_{M} / \hbar \right]                  \label{Zfac}
\EN
where the action  ${\cal S}_{M}$ is introduced in (\ref{Z}).  To obtain the
classical limit of the partition function we would relax the constraint
$x(0) = x(\beta \hbar)$ in (\ref{Zfac}):
\begin{eqnarray}
{\cal Z} \;=\; \int \frac{dxdp}{2 \pi \hbar} \; \int_{x(0) = x} {\cal D}
x(\tau) \;
\exp \left[ - \left( {\cal S}_{M} \,+\, ip(x(\beta \hbar) - x(0)) \right)
 / \hbar \right]          \nonumber \\
=\;   \int \frac{dxdp}{2 \pi \hbar} \; \int_{x(0\exp \left[  - \frac{1}{\hbar}
\int_{0}^{\beta \hbar} d\tau\,\left( \frac{1}{2}\,
\left(\dot{x}-ip \right)^{2}
\,+\, V(x) \,+\, \frac{1}{2}p^{2}  \right) \right]     \label{Zcons}
\end{eqnarray}
Eq. (\ref{Zcons}) allows for the smooth limit  $\hbar \rightarrow 0$ in which
the interval of $\tau$ vanishes. Then one ends up with a classical partition
function:
\EQ
{\cal Z}_{cl} \;=\; \int \frac{dxdp}{2 \pi \hbar} \;
\exp \left[  - \beta \left(  \frac{1}{2}p^{2}\,+\,V(x) \right) \right]
\label{Zcl}
\EN
An alternative way, suitable for $\hbar \neq 0$ is following \cite{dk} to
 expand
in $y(\tau)$ - fluctuations about the solutions $x_{E}(\tau)$ to the
Euclidean equations of motion with periodic boundary conditions $x_{E}(0) =
x_{E}(\beta \hbar) = x$. For the Green function (\ref{G}) one obtains:
\begin{eqnarray}
G_{\beta, \hbar}(\tau) \; \;=\; \; {\cal Z}^{-1} \;\int dx \, x\,
e^{-{\cal S}_{cl}}
\; \int {\cal D}y(\tau)  \; \left[ x_{E}(\tau) + y(\tau) \right] \nonumber \\
 \exp \left\{ - \frac{1}{\hbar} \int_{0}^{\beta \hbar}
d\tau \,\left[ y \left( \frac{-d^{2}}{2d \tau^{2}} \,+\, V^{''}(x) \right) y
 \,+\, W(x_{E},y) \right] \right\}                                \label{Gy}
\end{eqnarray}
where $W(x_{E},y)$ is $O(y^{3})$. The leading term, coming from the Gaussian
functional integral over $y(\tau)$ :
\EQ
G_{\beta, \hbar}(\tau) \; \;=\; \; {\cal Z}^{-1} \;\int dx \;
x\, x_{E}(\tau)\;
Det^{-1/2}(x) \; e^{-{\cal S}_{cl}(x)}                  \label{Gsemic}
\EN
depends on the determinant $Det$ of small fluctuations about $x_{E}(\tau)$
which is a
function of $x$ for  $\hbar \neq 0$. Note that the functional integral over
$y(\tau)$
is not saturated by smooth configurations and cannot be done semiclassically,
by definition. In the limit  $\hbar \rightarrow 0$ the integral over
 $y(\tau)$ is dominated by the "kinetic ebeing suppressed
due to the boundary condition $y(\beta \hbar)=y(0)=0$. So in that limit
the determinant $Det$ becomes equal to the integral over momentum $p$ in
 (\ref{Zcl}),
it does not depend on $x$ and cancels the corresponding integral in the
partition function:
\begin{eqnarray}
Det(\hbar \rightarrow 0) \;\;=\;\; \lim_{\hbar \rightarrow 0} \;
\int {\cal D}y(\tau)  \; \exp \left[  - \frac{1}{\hbar} \,\int_{0}^{\beta
\hbar} d\tau \; \frac{1}{2}\, \dot{y}^{2} \right]   \nonumber \\
=\;\; \lim_{\hbar \rightarrow 0} \; Tr \; e^{- \beta \, \hat{p}^{2} /2}
\;\;= \;\; \int \frac{dp}{2 \pi \hbar}\; e^{-\beta \, p^{2} /2}
\end{eqnarray}
 Thus in the classical limit Eq.(\ref{Gsemic}) becomes :
\EQ
G_{\hbar = 0}(\tau) \; \;=\; \; {\cal Z}_{st}^{-1} \;\int dx \, x\,
 x_{E}(\tau) e^{ - \beta V(x)}                  \label{Gstat}
\EN
where ${\cal Z}_{st}$ is a partition function of the static configurations :
${\cal Z}_{st} \, = \, \int dx \, \exp (- \beta V(x))$. We would like to
emphasize that the  $\hbar \rightarrow 0$ limit of $x_{E}(\tau)$ does exist,
although there is no interval to impose periodic boundary conditions any
more. $x_{E}(\tau)$ is a solution of Euclidean equations of motion with the
boundary condition $x_{E}(0) = x, \; \dot{x}_{E}(0) = 0$.

Let us turn to the evaluation of the classical 2-point Green function. In
accordance with (\ref{Zcons}), (\ref{Zcl}) it is defined as:
\EQ
G_{cl} \;\; =\;\;{\cal Z}_{cl}^{-1} \;\int \frac{dxdp}{2 \pi \hbar} \;\; x \;
x_{cl}(t,x,p) \; \; \exp \left[  - \beta H(p,x) \right]  \label{Gcl}
\EN
where $ x_{cl}(t,x,p)$ is a solution of the Hamiltonian equtions of motion
 ($t$ is real time) with the boundary conditions:
$x_{cl}(t=0)\,=\,x, \; \dot{x}_{cl}(t=0)\,=\,p$. The expression (\ref{Gcl})
was
evaluated exlicitly in the one-dimensonal case of the anharmoniin \cite{dk} and
it was found in strong disagreement with explicit expression
 for (\ref{Gstat}). Futher semiclassical corrections coming from (\ref{Gy}),
computed by Dolan and Kiskis did not eliminate that disagreement. We are now
in a position to clarify the nature of this discrepancy.

We already noticed that the Euclidean solution $x_{E}(\tau)$ survives in the
limit $\hbar \rightarrow 0$. The next observation is that an analytical
continuation $\tau = it$ of  $x_{E}(\tau,\, \hbar=0)$ will always give us a
 real time solution of the classical equations of motion $x_{cl}(t)$. To be
more precise:
\EQ
x_{E}(\tau=it,\, \hbar =0) \;\; =\;\;  x_{cl}(t,\,x,\,p=0)   \label{solutions}
\EN
Eq. (\ref{solutions}) implies a relation between the classical Green
function (\ref{Gcl}) and analytical continuation of the semiclassical
Matsubara calculations (\ref{Gy}). If one ignores the momentum dependence
$x_{cl}(t,x,p)$ in (\ref{Gcl}), the integration over $p$ in the numerator
cancels the corresponding term in the partition function (\ref{Zcl}), so one
gets precisely the analytical continuation of the leading term of the
Matsubara semiclassical expansion (\ref{Gstat}):
\EQ
G^{st}_{cl} \;\; =\;\;{\cal Z}_{st}^{-1}\;\int dx \;x \, x_{cl}(t,x,p=0) \;
e^{- \beta V (x)} \label{Gclst}
\EN
The static distribution in (\ref{Gclst}) is the one generated by the usual
first-order Langevin equation \cite{par}. It is the definition (\ref{Gclst})
that was used in the numerical calculations in real time in field theory at
finite temperature \cite{degra}, \cite{bdf}.

Expanding the solution $x_{cl}(t,x,p)$ in powers of momenta $p$ and
integrating over the distribution (\ref{Gcl}) yields a series of corrections
to the leading term (\ref{Gclst}). This expansion corresponds to the
semiclassical expansion (\ref{Gy}). It is important to realise that thexpansion
has purely geometrical character: it does not contain any small
parameter. The quantity in which we expand - $p$, is then being integrated
 over from $-\infty$ to $+\infty$. In this sense the expansion (\ref{Gy})
ineffecient.

We illustrate the statements above by explicit calculations in the simple
case of one degree of freedom with: $ H \; = \; p^{2}/2\,+\, V(x) \, , \;
 V(x) \; = \; \lambda \, x^{4}$. For the Euclidean solutions we have:
\EQ
x_{E}(\tau) \;\;=\;\; \frac{x_{o}}{{\em cn}\left[ 2 x_{o} \sqrt{\lambda}
(\tau\,-\,\beta \hbar /2 )\right]}                    \label{xE}
\EN
where $x_{o}$ is the turning point of vanishing velocity $\dot{x}$, determined
by $x$: $x_{o}/x\,=\, {\em cn}( x_{o} \sqrt{\lambda} \beta \hbar)$. ${\em cn}$
is Jacobian elliptic function of modulus $\kappa = 1/ \sqrt{2}$ \cite{book}.
The classical solution of the Hamiltonian equations of motion meanwhile is:
\EQ
x_{cl}(t) \;\;=\;\; x_{o}{\em cn}\left[ 2 x_{o} \sqrt{\lambda} t \;-\;
{\em cn}^{-1}(x/x_{o})      \right]        \label{xcl}
\EN
and the relation between $x$ and $x_{o}$ comes from the conservation of energy:
$\lambda x_{o}^{4}\;=\; p^{2}/2 \,+\,\lambda x^{4}$. Note that the difference
 between $x$ and $x_{o}$ is due to nonzero momentum $p$ in the classical case
and nonzero interval $\beta \hbar$ in the Matsubara calculations. The static
average (\ref{Gstat}) emerges, because $x_{o} \, \rightarrow \,x$, as either
$\beta \hbar$ or $p$ vanish. Eq.(\ref{Gstat}) holds in both classical and high
temperature limits.
In the limit  $\hbar \rightarrow 0$ one obviously obtains (\ref{xcl}) for
$p=0$ from  (\ref{xE}) , using the property of the elliptic functions
${\em cn}(it) \,=\, 1/{\em cn}(t)$ for $\kappa = 1/ \sqrt{2}$.

The difference between the exact classical Green function (\ref{Gcl}) and the
static approximation (\ref{Gclst}) is$x_{cl}(t,\,x,\,p)$ in powers of $p^{2}$
and integrating over $p$ the
$O(p^{2})$-term one gets:
\begin{eqnarray}
\Delta \, G_{cl} \;\;=\;\; \frac{1}{2 \Gamma(1/4) \sqrt{\lambda\beta}} \;
\int_{0}^{\infty} dy \;e^{-y^{4}} \nonumber \\
\left( \frac{{\em cn}(\gamma y) \, {\em sn}^{2}(\gamma y)}{y^{2}} \;+\;
\frac{1}{y} \, \frac{\partial}{\partial y} \, {\em cn}(\gamma y)  \right)
\end{eqnarray}
with $\gamma \,= \, 2 t (\lambda / \beta)^{1/4}$,
which is exactly the next to leading correction of the Euclidean semiclassical
expansion (\ref{Gy}), computed in \cite{dk}. Obviously the expansion
has no small parameter.

In conclusion, we have shown that the analytical continuation of the leading
semiclassical approximation to the temperature 2-point correlator in the
limit $\hbar \rightarrow 0$ is a static approximation to the classical Green
function, which is dealt with in the numercial simulations of the first-order
Langevin equation. The validity of our conclusions for many degrees of freedom
is rather evident, although the case of quantum field theory has a special
 feature that will be addressed in a separate publication.\\

{\large {\bf Acknowledgements}}

Fruitful discussions with R. Willey, P. Millard are gratefully acknowledged.
 I would like to thank D. Boyanovsky, R. Carlitz, A. Duncan for the discussion
of these results. This work was supported by the NSF grant PHY-9024764.

\newpage
{\large Figure Captions}\\

Fig. 1 a) Periodic Euclidean solutions $x_{E}(\tau)$ for two values of
$\beta \hbar$: $(\beta \hbar)_{2} \; > \; (\beta \hbar)_{1}$. b) Real time
trajectory $x_{cl}(t,x,p)$. Particle oscillates between the turning
points $-x_{o}$ and $+x_{o}$.


\begin{thebibliography}{99}

\bibitem{mats} T. Matsubara, {\em Progr. Theor. Phys. 14, 351 (1955)}.
\bibitem{mat} For a recent review see: M. P. Mattis, {\em Rhys. Rep. 214, 159
(1992)}.
\bibitem{rw} A. Ringwald, {\em Nucl. Phys. B330, 1 (1990)};\\
O. Espinosa, {\em Nucl. Phys. B334, 310 (1990)}.
\bibitem{dk} L. Dolan, J.Kiskis, {\em Phys. Rev. D20, 505 (1979)}.
\bibitem{par} G. Parisi, Wu Yongshi, {\em Scientia Sinica 24\bibitem{degra} R.
Loft, T. DeGrand, {\em Phys. Rev. B35, 8528 (1987)}.
\bibitem{bdf} A. Bochkarev, Ph. de Forcrand, {\em Phys. Rev. Lett. 63, 2337
(1989)}; {\em Phys. Rev. D44, 519 (1991)}.
\bibitem{book} Our notations follow P.F. Byrd, M.D. Friedman, {\em
Handbook of Elliptic Intergrals for Engineers and Scientists}
(Springer, New York, 1971).

\end{thebibliography}
\end{document}